\documentclass[twocolumn,nofootinbib]{revtex4-1}
\usepackage{amsmath}
\usepackage{amsthm}
\theoremstyle{definition}

\usepackage{xcolor}
\usepackage{tensor}
\usepackage{soul}
\usepackage{hyperref}
\newcommand{\eqnref}[1]{Eq.~(\ref{#1})}

\usepackage{amssymb}

\newcommand{\appendixref}[1]{Appendix \ref{#1}}

\newcommand{\myparagraph}[1]{\section{#1}}

\begin{document}
%TC:ignore
\title{Strange metals as ersatz Fermi liquids}
\author{Dominic V. Else}
\affiliation{Department of Physics, Massachusetts Institute of Technology, Cambridge, MA 02139, USA}
 \author{T. Senthil}
\affiliation{Department of Physics, Massachusetts Institute of Technology, Cambridge, MA 02139, USA}
\begin{abstract}
A long standing mystery of fundamental importance in correlated electron physics is to understand strange non-Fermi liquid metals that are seen in diverse quantum materials. A striking experimental feature of these metals is a resistivity that is linear in temperature ($T$). In this paper we ask what it takes to obtain such non-Fermi liquid physics down to zero temperature in a translation invariant metal. If in addition the  full  frequency ($\omega$) dependent conductivity satisfies $\omega/T$ scaling,  we argue that the $T$-linear resistivity must come from the intrinsic physics of the low energy fixed point.   Combining with earlier arguments that compressible translation invariant metals are `ersatz Fermi liquids' with an infinite number of emergent conserved quantities, we obtain powerful and practical conclusions. We show that there is necessarily a diverging susceptibility for an operator that is odd under inversion/time reversal symmetries, and has zero crystal momentum.   We discuss a few  other experimental consequences  of our arguments, as well as potential loopholes which necessarily imply other exotic phenomena. 
\end{abstract}

\maketitle

%TC:endignore

A diverse variety of `strange' metals are seen \cite{lee2006doping,proust2019remarkable,varma2020colloquium,lohneysen2007fermi,gegenwart2008quantum}
to not fit
the basic predictions of Fermi liquid theory. Examples include cuprate high temperature superconductors, some heavy electron materials tuned to a quantum critical points, and a growing number of other correlated metals (see, eg, Refs. \cite{shibauchi2014quantum,oike2015pressure,doiron2009correlation,cao2020strange} )
The non-Fermi liquid physics manifests itself through unconventional power laws that go down to energy scales much lower than any microscopic scale.  A striking example is a resistivity that increases linearly with temperature over a wide range that extends to very low temperature. There is currently very little understanding of this linear resistivity and other properties in most experimental systems.  Understanding these strange metals is widely regarded as one of the biggest challenges in modern physics. 

 Here we present a number of general theoretical observations that provide strong restrictions on the dynamics of a class of {\em clean} strange metals. We expect that this class includes both the cuprate strange metal as well as non-Fermi liquid heavy fermion quantum critical metals. Remarkably we show, under some very general conditions discussed below, that obtaining a linear  resistivity down to $T = 0$ in a clean metal requires the divergent susceptibility of an observable that is odd under inversion/time reversal, transforms as a vector under lattice rotations,  and has zero crystal momentum. These are the same symmetries as those of the 
loop current order parameter\cite{simon2002detection} discussed in the cuprate materials. Thus our discussion of strange metal transport provides a very general reason for a diverging loop current  susceptibility in the strange metal  which may connect to the various reports and controversies (for a sampling of some representative papers, see Refs. \cite{fauque2006magnetic,mook2008observation,li2008unusual,xia2008polar,zhao2017global,lederer2012observable,mounce2013absence,wu2015incipient,croft2017no,bourges2018comment,zhang2018discovery,gheidi2020absence}) surrounding such order in the proximate pseudogap metal. 

We consider a putative non-Fermi liquid metal with  the following assumed properties:
\begin{enumerate}
\item \textit{Clean}. \label{assump:clean} The system microscopically has $\mathrm{U}(1)$ charge conservation symmetry and lattice translation symmetry (no disorder).

\item \textit{Conductivity scaling}. \label{assump:scaling}
At low temperatures and frequencies, the conductivity approaches the universal scaling form
\begin{equation}
\label{eq:conductivity_scaling}
 \sigma(\omega,T) = T^{-1} \Sigma(\omega/T)
\end{equation}
for some function $\Sigma$, such that $\Sigma(0)$ is a nonzero finite number. In particular, the DC resistivity is proportional to $T$.
\item \textit{Compressible}.
\label{assump:compressible}
The charge $\nu$ per unit cell can be continuously tuned as a function of some microscopic parameters without affecting the above properties, and is not pinned to any particular rational value.
\end{enumerate}

These assumptions are strongly motivated by the observed non-Fermi liquid physics in the cuprates and at heavy electron quantum critical points. We could perhaps refer to these assumptions as our ``Central Dogmas'' \cite{Crick,Anderson}.  
So 
let us briefly discuss the experimental evidence for these assumptions.  

We begin with the first assumption, whose non-trivial content is that the observed behavior is a property of a clean lattice system.  Real materials of course have some level of disorder that breaks lattice translation invariance. Our assumption then is that to understand the essence of the strange metal physics including the linear-$T$ resistivity, the disorder is unimportant. 
Support for this assumption comes from studies\cite{rullier2000universal} on cuprates that are artificially damaged by electron irradiation, which provides a gentle way of tuning the disorder strength. It is seen that such irradiation increases the residual resistivity at zero temperature but does not change the slope of the linear resistivity. Moreover, for the cleanest samples the residual resistivity can be made very small, and generally is much smaller than the total resistivity in most of the temperature range where linear resistivity is observed. This suggests that the residual resistivity is related to disorder while the physics of linear resistivity is not affected by such disorder, and that there is a hypothetical perfectly clean limit in which the residual resistivity goes to zero while the linear resistivity remains.
 For heavy fermion quantum critical metals, some of them like $\mathrm{YbRh}_2\mathrm{Si}_2$ are stoichiometric compounds and  it is perhaps not unreasonable that the basic  non-Fermi liquid physics is not determined by disorder effects. 
 
Next we consider the second assumption, of conductivity scaling. Linear dc resistivity down to ultra-low temperatures is of course seen in many non-Fermi liquid metals. We note, however, that our assumption implies the absence of a residual zero temperature resistivity; as mentioned above, we expect that in the clean limit, the residual resistivity would indeed go to zero. The $\omega/T$ scaling of the frequency dependent conductivity has been directly demonstrated recently\cite{Prochaska285} in $\mathrm{YbRh}_2\mathrm{Si}_2$. Evidence for such scaling in the cuprate strange metal regime has long been reported\cite{van2003quantum}, at least upto $\hbar \omega$ slightly bigger\footnote{At higher frequency this scaling is obscured by a power law conductivity which is possibly of a different origin from the putative quantum criticality; this high frequency power law extends upto an energy of order $1 eV$ comparable to microscopic scales.} than $k_BT$.  

Finally the third assumption - that the metal is compressible - is  widely made in the literature though it has not been scrutinized in detail experimentally. In the cuprates the hypothesized quantum critical doping associated with the strange metal occurs at slightly different values in different materials. This is consistent with assuming that the critical doping can be continuously tuned by varying microscopic parameters. It may be possible to demonstrate this directly by studying the change of critical doping with pressure in a single cuprate material.

We will  obtain some striking theoretical constraints on metals with these assumed properties, without actually constructing  a specific model of such a metal. Assumptions \ref{assump:clean} and \ref{assump:compressible} are also shared by conventional Fermi liquids in clean systems; however, these manifestly do not satisfy Assumption \ref{assump:scaling}. For example, the DC conductivity of a clean Fermi liquid scales like $\sigma(0,T) \propto T^{-2}$ (or faster if umklapp is not effective). 
Our discussion builds on the results of Ref. ~\cite{Else_2007} which focused on  the kinematics of compressible translation invariant quantum phases/phase transitions.  A key result  is that any such metallic phase has a very large emergent symmetry and associated conservation laws.  Non-Fermi liquids with such an emergent symmetry were dubbed `ersatz Fermi liquids'. Here we focus  on the dynamics of such ersatz Fermi liquids. 

\myparagraph{Strange metal transport is  ``intrinsic''}
An important distinction to make is between ``intrinsic'' and ``extrinsic'' resistivity. It is helpful to use  the language of the renormalization group (RG). Quite generally  the low-energy physics of the system is described by some RG fixed point. The resistivity is ``intrinsic'' if this RG fixed point theory itself has nonzero DC resistivity at nonzero temperature. By contrast, the resistivity is ``extrinsic'' if the DC resistivity of the RG fixed point theory is zero (even at nonzero temperature); then nonzero resistivity must arise entirely from RG-irrelevant couplings.

In a conventional clean Fermi liquid, the resistivity is extrinsic. There the only source of resistivity is umklapp scattering which  is an irrelevant perturbation to the Fermi liquid fixed point\footnote{There is a slight subtlety here. Within the framework of the usual Shankar RG\cite{shankar1994renormalization}, Ref. \cite{kumar1996umklapp} showed that the umklapp terms were actually formally marginal but nevertheless did not affect the physics in the low energy limit. when the width of the momentum window $\Lambda$ where the theory is defined is taken to zero at a fixed Fermi momentum.   We expect that a slightly different formulation of the RG, which keeps track of the running of the Fermi momentum, will render the umklapp terms formally irrelevant in agreement with their unimportance for the physics.
}.
%However, we  know  that Fermi liquids do not satisfy the strange metal conductivity scaling (Assumption \ref{assump:scaling}). 
By contrast, for systems satisfying Assumption \ref{assump:scaling}, the resistivity must be intrinsic. To see this, note that the conductivity of the system as a function of frequency, $\sigma(\omega,T)$, in general depends on the values of irrelevant couplings. However, the only way for the asymptotic behavior at small $\omega$ and $T$ to be described by \eqnref{eq:conductivity_scaling}, is if the right-hand side of \eqnref{eq:conductivity_scaling} represents the conductivity \emph{of the fixed point theory} with irrelevant terms set to zero; we give a careful proof of this statement in \appendixref{appendix:intrinsic_proof}. Since Assumption 2 then states that $\Sigma(0) < \infty$, it follows that the DC conductivity of the fixed-point theory is not infinite. We remark that at first glance \eqnref{eq:conductivity_scaling} as a result for an RG fixed-point theory might seem surprising from the point of view of dimensional analysis. However, recall that the Fermi liquid fixed point also satisfies \eqnref{eq:conductivity_scaling}, albeit with the scaling function $\Sigma$ being a delta function. The point is that in the RG for the Fermi liquid fixed point, there is a length scale $k_F$ (the Fermi wave-vector) that does not scale in the RG flow.

\myparagraph{The emergent symmetries of a strange metal}
 Now we turn to examining the consequences of the assumptions that the metal is clean, and is compressible (Assumptions \ref{assump:clean} and \ref{assump:compressible}). We first recall the result of  Ref.~\cite{Else_2007}. For any system satisfying Assumptions \ref{assump:clean} and \ref{assump:compressible}, the group of emergent internal symmetries in the IR fixed point theory \emph{cannot} be a compact finite-dimensional Lie group. 
What then could the emergent internal symmetry group of a strange metal be?  A hint is provided by  ordinary Fermi liquids which, since they satisfy Assumptions \ref{assump:clean} and \ref{assump:compressible} must indeed obey the constraints of Ref.~\cite{Else_2007}.

A Fermi liquid manages to have a symmetry group that is not a compact finite-dimensional Lie group because that the charge at \emph{each} point on the Fermi surface is separately conserved. Specifically (in two dimensions, say) any operator of the form
\begin{equation}
\int f(\theta) \hat{n}(\theta) d\theta
\end{equation}
is conserved, for any smooth function $f(\theta)$, where $\theta$ is some coordinate parameterizing the Fermi surface, and where $\hat{n}(\theta)$ is the linear charge density operator with respect to $\theta$; thus the total charge of the system is $\hat{Q} = \int \hat{n}(\theta) d\theta$, while the momentum can be expressed as
\begin{equation}
\label{eq:momentum}
\hat{\mathbf{P}} = \int \mathbf{k}(\theta) \hat{n}(\theta) d\theta,
\end{equation}
where $\mathbf{k}(\theta)$ is the momentum of point $\theta$ on the Fermi surface. Therefore, the emergent internal symmetry group of the Fermi liquid is an \emph{infinite}-dimensional continuous group. As explained in Ref. \cite{Else_2007} this emergent symmetry is anomalous. Some (though not all) universal properties of a Fermi liquid can be understood directly in terms of its emergent symmetry and the associated anomaly. 

It is interesting to  postulate that the emergent internal symmetry and anomaly of the strange metal fixed point is  the same as a Fermi liquid, despite having no quasiparticles.  Ref.~\cite{Else_2007} introduced the term \emph{ersatz Fermi liquid} to refer to such a system,
Now we will examine the very striking consequences of this postulate. In Section \ref{appendix:loopholes}, we discuss other possibilities.

\myparagraph{Charge transport in ersatz Fermi liquids}
%\label{sec:charge_transport}
There is a tension between the strange metal being an ersatz Fermi liquid and having nonzero resistivity. Any conserved quantity risks leading to dissipationless current flow if it has nonzero overlap with the electrical current, since the conservation law then prevents the current from fully relaxing.

For simplicity, let us first consider the case of an ersatz Fermi liquid in two spatial dimensions with continuous rotational symmetry. In that case, the only conserved quantities that can overlap with the current are $n_1$ and $n_{-1} = n_1^{\dagger}$, where we defined the Fourier components of the $\hat{n}(\theta)$'s: $\hat{n}_l = \frac{1}{2\pi} \int_0^{2\pi} e^{-il\theta} \hat{n}(\theta) d\theta$. These are closely related to the momentum, since we have $\mathbf{k}(\theta) = k_F (\cos \theta, \sin \theta)$ which implies from \eqnref{eq:momentum} that $P_x = \pi(\hat{n}_1 + \hat{n}_{-1})$ and $P_y = \pi(\hat{n}_1 - \hat{n}_{-1})/i$.  Then \cite{Forster,Hartnoll_1201,Lucas_1502} (see also \appendixref{appendix:susceptibilities} for an easy argument) the real part of the frequency-dependent conductivity is given by
\begin{equation}
\label{eq:sigmaomega}
\sigma(\omega) = \frac{\pi \mathcal{Q}^2}{\mathcal{M}} \delta(\omega) + \mbox{(non-singular part)},
\end{equation}
where $\mathcal{Q}$ is the charge density, and
\begin{equation}
\label{eq:Mdefn}
\mathcal{M} := \frac{1}{V} \chi_{P_x P_x} := \frac{1}{V} \left( \frac{\partial}{\partial v} \right) \langle P_x \rangle_{H - v P_x} \bigg|_{v=0},
\end{equation}
can be interpreted as the ``mass density'', where $\langle \cdot \rangle_{H - v P_x}$ denotes a thermal expectation value with the Hamiltonian $H$ replaced by $H - v P_x$.
The delta function in \eqnref{eq:sigmaomega} leads to infinite DC conductivity (unless its coefficient is zero). This is an example of the ``momentum bottleneck'' for current relaxation.
%It is due to the momentum bottleneck that models attempting to describe the strange metal based on, for example,  holography tend to include explicit momentum relaxation (see Ref.~\cite{HolographicQuantumMatter} for a review).   However, in this paper we have been inexorably led, as described previously, to the conclusion that the $\hat{n}(\theta)$'s are conserved in the IR theory and therefore we cannot contemplate any symmetry-breaking terms.

Is there any way to suppress the delta function in \eqnref{eq:sigmaomega} in order to obtain finite DC conductivity? Strange metals are supposed to exist at finite charge density, so $\mathcal{Q} \neq 0$. Therefore, the only way to suppress the delta function is if \emph{$\mathcal{M}$ is infinite}. 
Going beyond the assumption of continuous rotational symmetry, and for any spatial dimension, we show in \appendixref{appendix:susceptibilities} that it remains the case that the only way to suppress the delta function in the conductivity at zero frequency for an ersatz Fermi liquid, assuming generic charge density [Assumption \ref{assump:compressible}], is for a certain susceptibility of the $\hat{n}(\theta)$'s to diverge. Therefore, we have reached one of the principal conclusions of our paper: Assumptions \ref{assump:clean}--\ref{assump:compressible}, if satisfied by way of the system being an ersatz Fermi liquid, imply the divergence of a susceptibility of the emergent conserved quantities.

 Note that, as defined by \eqnref{eq:Mdefn}, $\mathcal{M}$ is the suceptibility of a quantity $\hat{P}_x$ that is odd under time-reversal and inversion symmetry.   In fact, in \appendixref{appendix:susceptibilities}  we show that even without continuous rotation symmetry, and in any spatial dimension, the operator for which the divergent susceptibility suppresses the delta function in \eqnref{eq:sigmaomega} must share the same symmetry properties as the electrical current operator. Thus, it is odd under time-reversal symmetry and inversion symmetry, while under  lattice rotation symmetry, it transforms as a vector. 
 This suggests that the divergent susceptibility could potentially be a signature of a continuous phase transition into a phase that (among other features) spontaneously breaks inversion and time-reversal symmetry, a point we return to later.  

Finally, note that at any $T > 0$ the susceptibility will probably be finite, while the emergent conservation laws will be violated by irrelevant operators. Since the actual microscopic system always contains irrelevant couplings, at nonzero temperature there will be an interplay between two effects: firstly, one expects (say in the continuous rotation symmetry case) that $\mathcal{M}$ becomes finite; secondly the conserved quantities $\hat{n}(\theta)$ are no longer perfectly conserved and instead acquire a finite lifetime $\tau$. In that case, one expects the delta function in \eqnref{eq:sigmaomega} to be replaced by a Lorentzian with width $\tau^{-1}$.
This gives a coherent contribution to the DC conductivity, namely \newcommand{\sdcc}{\sigma_{\mathrm{DC}}^{\mathrm{coherent}}}
$\sdcc = \mathcal{Q}^2 \tau / \mathcal{M}$. The scaling of $\sdcc$ as the temperature goes to zero will then depend on a competition between how fast $\mathcal{M}$ and $\tau$ diverge. It is possible that $\sdcc$ could fail to go to zero, or even diverge, as $T \to 0$. 

One might object that if $\sigma_{\mathrm{DC}}^{\mathrm{coherent}}$ does not go to zero as $T \to 0$, then our Assumption 2 is not satisfied, and hence the arguments of our paper are not applicable in the first place. However, if one formulates our Assumption \ref{assump:scaling} in a sufficiently precise way in terms of distributions as we do in Appendix \ref{subsec:intrinsic_precise}, then one can show that this formulation of the assumption is still satisfied provided that $\tau$ diverges more rapidly than $\sim 1/T$ as $T \to 0$, as it should since the finite lifetime is due to an irrelevant coupling.

Note that in practice, one does not necessarily expect to experimentally observe the $\sdcc$ contribution to the DC conductivity; because the weight $\mathcal{Q}^2/\mathcal{M}$ goes to zero as $T \to 0$, at low temperatures even a small amount of disorder could lead to a large but finite $\tau$ and thus suppress $\sdcc$.

\section{Experimental tests}

We next describe a number of  experimental tests of the idea that the strange metal is an ersatz Fermi liquid.  

\subsection{Crossover to off-critical resistivity and scaling}
One signature  is the scaling of resistivity  in cases where the strange metal occurs at a quantum critical point proximate to a Landau Fermi liquid.  
%Suppose that we are in the ersatz Fermi liquid, and tune some parameter such as doping in order to move away from the strange metal, and thereby reach a Landau Fermi Liquid ground state. 
 Then, away from criticality,  $\mathcal{M}$ in \eqnref{eq:sigmaomega}will become finite, thereby reactivating  the mechanism of conserved quantities preventing current decay. The conductivity will be dominated at low frequencies and temperatures by the delta function peak in \eqnref{eq:sigmaomega} (which can get broadened with width $\propto T^2$ due to momentum relaxation from irrelevant couplings). But since the weight of this peak precisely goes to zero at the critical point, where the conductivity must instead have a different origin, we should \emph{not} expect the conductivity at low temperatures and frequencies near the critical point to collapse onto a universal scaling curve.  By contrast, if  one of the loopholes discussed in Section \ref{appendix:loopholes} applies, and the strange metal is not an ersatz Fermi liquid, then it is conceivable that such a scaling collapse could occur. 
%In particular if the internal symmetry group of the low energy theory is $U(1) \times Z^d$ that we discussed earlier, the scaling out of criticality into the Fermi liquid will naturally incorporate momentum relaxation; then we expect that a single scaling function will capture both the quantum critical and Fermi liquid resistivities near the critical point. 
This  conclusion will hold irrespective of the nature of the detailed crossover from the ersatz Fermi liquid quantum critical fixed point to the ordinary Fermi liquid. 

 It is sometimes observed that in the proximate Fermi liquid near a strange metal quantum critical point, the resistivity $\rho(T) -\rho(0) = AT^2$ with $A$ diverging upon approaching the critical point\footnote{This is seen in heavy fermion quantum critical points. The situation in the cuprates is murky. An analysis of overdoped-$\mathrm{La}_{2-x} \mathrm{Sr}_x \mathrm{CuO}_4$ (LSCO)  in a high magnetic field\cite{cooper2009anomalous} surprisingly found that the resistivity was linear at low $T$ through out the overdoped region. However this analysis involved exrapolating the measured high field resistivity to zero field by subtracting a quadratic magnetoresistance. The data (at least in the vicinity of optimal doping) is actually better fit by a linear magnetoresistance in the measured field range ; indeed recent experiments\cite{giraldo2018scale}  on LSCO thin films find such a linear magnetoresistance.  Thus a clear understanding of whether the low-$T$ overdoped cuprate is a Fermi liquid or not awaits clarification. } while the critical point itself shows linear resistivity. For an ersatz Fermi liquid, the diverging $A$ coefficient {\em should not} be part of a scaling function with the linear resistivity. Such a scaling has been attempted in heavy electron systems (Ref.~ \cite{custers2003break}); our discussion calls for careful scrutiny of this scaling plot.  It would also be very interesting to determine how exactly the AC conductivity crosses over from the ``broad'' peak of width $\sim T$ in the strange metal [see \eqnref{eq:conductivity_scaling}] to the much narrower Drude peak in the Fermi liquid.

%The  iron pnictide material $\mathrm{BaFe}_2 ((\mathrm{As})_{1-x}\mathrm{P}_x)_2$  shows a broad regime of linear-$T$ resistivity that has been associated with a quantum critical point. Refs. \cite{hayes2016scaling,hayes2018magnetoresistance}  show that in a magnetic field  the resistivity satisfies the scaling form $\rho(H, T) - \rho(0,0) = \sqrt{ a T^2 + c H^2}$ which clearly crossover over to a Fermi liquid-like $T^2$ resistivity in a non-zero field at the lowest temperatures. This scaling may suggest that the putative quantum critical metal in this system is not an ersatz Fermi Liquid; however the role of disorder, and the lack of momentum conservation induced by the magnetic field  must be considered carefully before we can reach firm conclusions. Another way for this result to remain consistent with an ersatz Fermi liquid  would be that the magnetic field, being a very special perturbation, might not tune the mass density $\mathcal{M}$ away from infinity at leading order. Note that a scaling like that of Refs.~\cite{hayes2016scaling,hayes2018magnetoresistance} does not appear to hold in the cuprate $\mathrm{Bi}$-2212 \cite{legros2019universal}.
%  

\subsection{Inversion/time reversal breaking order}
%\label{subsec:inversion_breaking}
We argued - on general grounds - that if the strange metal is an ersatz Fermi liquid (or a variant) then it is necessary that a susceptibility of the emergent conserved $\hat{n}(\theta)$ must diverge in order to obtain the required resistivity. For this to work, the susceptibility must diverge in channels that have overlap with the current operator.  Thus the low energy theory has observables $O$ which are time reversal/inversion odd, live at zero crystal momentum, transforms as a vector under lattice rotations, and whose susceptibility diverges.  This is a firm prediction of the ersatz Fermi liquid hypothesis for strange metals that could potentially be tested. 

In the cuprates,  there have been many reports (and  controversies) of ordering that spontaneously breaks precisely the symmetries of such observables $O$ (see, eg, Refs.~ \cite{fauque2006magnetic,mook2008observation,li2008unusual,xia2008polar,zhao2017global,lederer2012observable,mounce2013absence,croft2017no,bourges2018comment,zhang2018discovery,gheidi2020absence}). These have been usually interpreted microscopically in terms of loop current ordering.  Remarkably, our considerations, which come from a completely different line of thought, demand the existence of critically diverging  fluctuations of such order in the strange metal regime.  This may be consistent with the emergence of static order in the pseudogap ground state. However we caution that the pseudogap ground state is not {\em just}  an ordinary  Fermi liquid metal in the presence of such order.  Rather on top of whatever transformation underlies the evolution between the overdoped and underdoped metallic ground states (eg, a Fermi surface jump), our considerations make it plausible that there is a breaking of time reversal/inversion symmetries. Furthermore the diverging susceptibility of an order parameter at a quantum critical points does not necessarily imply that one of the proximate phases has static order for the corresponding observable, as is known from a number of theoretical examples.

\subsection{Quantum oscillations}
Another experimental test that one could consider in principle, although in practice it may be difficult to realize, is based on quantum oscillations.  Consider any system in two spatial dimensions with lattice translation symmetry and $\mathrm{U}(1)$ charge conservation symmetry, and let $\nu$ be the average charge per unit cell. Then we say the system exhibits \emph{universal quantum oscillations} if, upon applying a weak magnetic field $B$, the properties of the system (for example, resistivity) are periodic in $1/B$ with period
\begin{equation}
\Delta (1/B) = \frac{e}{2\pi \hbar} \frac{1}{\rho}
\end{equation}
where $\rho$ is some number (the ``effective charge density'') such that $\rho V_{\mathrm{unit}} = \nu \, [\mathrm{mod} \, 1]$, where $V_{\mathrm{unit}}$ is the volume of the unit cell. (For spinful systems, there is an additional factor of 2 in this relation).
 
 Quantum oscillations were originally derived for Fermi liquids, based on a semiclassical quantization argument for the orbits of quasiparticles. However, in fact one expects quantum oscillations for any system where the discrete microscopic translation symmetry gets extended to an emergent continuous  symmetry\footnote{To be more precise, we say for a system in $d$ spatial dimensions that the microscopic translation symmetry gets extended to an emergent continuous symmetry if the emergent symmetry of the low-energy theory has a subgroup $G_{\mathrm{trans}} \cong \mathbb{R}^d$ such that the microscopic translation symmetry acts on the low-energy theory through a $\mathbb{Z}^d$ subgroup of $G_{\mathrm{trans}}$. Note that $G_{\mathrm{trans}}$ could, and will except in the case of ultra-local quantum criticality  (see Section \ref{subsec:ultralocal}), represent an internal symmetry.} \cite{Else_2007}, as happens for example for an ersatz Fermi liquid in which the Fermi surface does not wrap non-trivially around the Brillouin zone. We expect, moreover, that the converse also holds, so that universal quantum oscillations can be considered an experimental signature of the microscopic translation symmetry getting extended to an emergent continuous symmetry. When this occurs, then by a similar argument to the ersatz Fermi liquid case discussed above but, we emphasize, \emph{without} needing to assume the system is an ersatz Fermi liquid, then the only way to get intrinsic resistivity would be to have a diverging ``mass density'', corresponding to a diverging susceptibility.
 
Unfortunately, for hole-doped cuprate materials that are clean enough that one can expect to observe quantum oscillations, the critical magnetic field required to suppress superconductivity down to zero temperature at critical doping is larger than is accessible with current technology; for example it is estimated that the critical field is about 150 T in YBCO \cite{Grissonnanche_1303}.
  Meanwhile, at heavy fermion critical points, a magnetic field tunes the system out of criticality, again complicating a direct determination of the possibility of quantum oscillations associated with the quantum critical state.
 Therefore, it has not been possible to verify whether the zero-temperature quantum critical point that is believed to control a strange metal exhibits quantum oscillations. However, hopefully this might be possible in the future.
 
 We remark that \emph{electron}-doped cuprates 
 exhibit $T$-linear resistivity in a range of dopings \cite{Fournier_1998,Jin_1108} (see Refs.~\onlinecite{Armitage_0906,Greene_1905} for reviews); quantum oscillations have been reported for some of these materials \cite{Higgins_1804} but not in the same doping range as the $T$-linear resistivity.
 This seems worthy of further study.

\section{Potential loopholes}
\label{appendix:loopholes}
The discussion in this paper has been based on the assumption that the strange metal is an ersatz Fermi liquid. It is important to consider what other possibilities are available (we confine ourselves to spatial dimensions $d \geq 2$ where the results of Ref.~\cite{Else_2007} should apply).
Let us remark from the outset, however, that all of the loopholes described below might be difficult to reconcile with the experimental observation in cuprates of a sharp Fermi surface that seemingly obeys Luttinger's theorem. As discussed in Ref.~\cite{Else_2007}, all ersatz Fermi liquids share these properties, but it is not clear how they would arise more generally in the absence of a conserved charge associated with each point in the Fermi surface, as occurs in an ersatz Fermi liquid.

\subsection{Ultra-local quantum criticality}
\label{subsec:ultralocal}

One possibility is that the arguments of Ref.~\cite{Else_2007} simply do not hold. Of course, these arguments were never mathematically rigorous, but at the physical level it is hard to see what could go wrong. There is only one possible loophole that we can think of, as follows. A central assumption of Ref.~\cite{Else_2007} was that the microscopic lattice translation symmetry becomes an internal symmetry in the IR theory. The rationale for this is that the renormalization group flow should involve a spatial rescaling, and therefore the effective unit cell size goes to zero. The loophole would be if there is \emph{no spatial rescaling} in the renormalization group flow (only time rescaling) at any point in the flow starting from the microscopic system.  Physically this means that the spatial correlation length of the critical fluctuations  is finite while the correlations in time are long ranged. This seems like an incredibly strong condition and it is not clear whether it could ever be satisfied in a clean many body system which microscopically has a  finite number of degrees of freedom per unit cell. We refer to this condition as ``Ultra-local quantum criticality". 

We emphasize the distinction with popular notions of `local quantum criticality' which figure prominently in many discussions of strange metals.  In contrast to what we discussed in the previous paragraph the usual  local quantum criticality  scenarios do {\em not} present  a loophole  to the results of Ref. \cite{Else_2007}. One version of local quantum criticality is where the spatial correlation length  diverge but only slower than any power of the time correlation length. In clean systems such criticality has been discussed in, eg, Ref.~\cite{zhu2015local} in dissipative $XY$ models; generalizations have been presented as a possible theory of the cuprate strange metal - for a review see Ref.~\cite{varma2020colloquium}. Both this version of local quantum criticality and the ``Ultra-local quantum criticality" may be loosely characterized by saying that the dynamical critical exponent $z = \infty$. However evading the constraints of Ref.~\cite{Else_2007} requires not just that $z = \infty$ but that the spatial correlation length stays finite. 

 A different version of local quantum criticality  - popular in the literature\cite{si2001locally} on  heavy fermion criticality - is where both spatial and temporal correlations diverge as power laws but only the temporal correlations have an anomalous exponent.  The corresponding dynamical critical exponent $z$ is finite. This scenario too, if realized, will be subject to the constraints of Ref.~\cite{Else_2007}.

Note that certain holographic theories of non-Fermi liquids \cite{Chamblin_9902, Liu_0903, Faulkner_1001,Gubser_0911,Davison_1311} have $z = \infty$, which probably means that they can evade the result of Ref.~\cite{Else_2007}, at least at the level of the holographic theories themselves, if one treats them as an effective theory and does not worry about their origin from the microscopic degrees of freedom.
However, from the point of view of a microscopic lattice model, these theories can at best represent the IR limit.
If there were such a lattice model, it would have to exhibit ultra-local quantum criticality to avoid generating an internal symmetry. However, it is not clear that such a microscopic lattice model exists.

\subsection{Non-compact finite dimensional symmetry group}

Another possibility is that the the emergent internal symmetry group is a finite-dimensional Lie group, but is non-compact. This is in principle allowed by the results of Ref.~\cite{Else_2007}, but a non-compact finite-dimensional internal symmetry group is a somewhat bizarre thing to contemplate in a theory that arises from a microscopic lattice model with finite Hilbert space dimension per site, and we do not know of any examples that could arise in condensed matter systems\footnote{Note that the infinte-dimensional emergent symmetry group of an ersatz Fermi liquid, the ``loop group'' $\mathrm{LU}(1)$ \cite{Else_2007}, while it is non-compact in the literal sense, should still perhaps be thought of as an infinite-dimensional generalization of finite-dimensional compact Lie groups, rather than of finite-dimensional non-compact Lie groups, since it shares a number of properties with the former.}.
Still, since all the alternatives are also highly exotic, we must still keep this possibility in mind.

If such an emergent symmetry is allowed then we will not have an infinite number of emergent conserved quantities. It is interesting to contemplate a simple example (in $d$ spatial dimensions) where the emergent internal symmetry is $U(1) \times Z^d$. Then microscopic lattice translations embed in the low energy theory as the discrete internal symmetry group $Z^d$. 
In this case, in the low energy theory the only conserved quantity corresponding to a non-discrete symmetry  (it is only these which potentially lead to infinite conductivity) in the low energy theory is the total charge itself. Then we do not have to worry about any mixing of the current with momentum (or other conserved quantities). A theory with this emergent symmetry will have an intrinsic conductivity which may match Eqn. \ref{eq:conductivity_scaling}.  

\subsection{Considering other possibilities}
Next, we could imagine that the emergent symmetry group is indeed an infinite-dimensional group, but different from that of an ersatz Fermi liquid. However, it seems very likely that the discussion above and in Appendix \ref{appendix:susceptibilities} would generalize to such a case, and one would again conclude that there must be a diverging susceptibility of the conserved quantities.

A final possibility is that the system could have emergent higher-form symmetries, which also are a potential loophole in the results of Ref.~\cite{Else_2007}. However, as discussed there it does not seem likely that such higher-form symmetries could allow the system to exist at generic charge filling $\nu$, unless there is also an infinite-dimensional 0-form symmetry group present, in which case the arguments of this paper would still apply.

\myparagraph{Conclusion}
In this work, we have unveiled a powerful new approach to understanding strange metals. Rather than trying to find theoretical models that can reproduce the phenomenology, which has so far eluded the community, we are able to make considerable progress through general structural arguments based only on minimal assumptions. Through such an approach, we have made strong model-independent predictions about the nature of strange metals. We expect that our results will  narrow down the search for a theoretical understanding  of these mysterious and fascinating phases of matter. 

%TC:ignore

\begin{acknowledgments}
We thank Sean Hartnoll, Andrew Lucas, Subir Sachdev, Ryan Thorngren, Shivaji Sondhi, Steve Kivelson, Chetan Nayak, and Jan Zaanen for helpful discussions and comments. We thank Zhengyan Shi for pointing out an error in an equation.
D.V.E.\ was supported by the EPiQS Initiative of the Gordon and Betty Moore Foundation, Grant No. GBMF8684. T.S.\ is supported by a US Department of Energy grant DE- SC0008739, and in part by a Simons Investigator award from the Simons Foundation. This work was also partly supported by
the Simons Collaboration on Ultra-Quantum Matter, which is a grant from the Simons Foundation (651440, TS).
\end{acknowledgments}

%TC:endignore

\appendix

\section{Proof of intrinsic resistivity}
\label{appendix:intrinsic_proof}

In this section, we will give the argument that our Assumption 2 implies that the resistivity must be intrinsic; that is, the DC conductivity of the fixed point theory must be finite. For clarity, we first give a quick, not totally precise argument, followed by a more precise version of the argument.

\subsubsection{Rough argument}
 In general  the DC conductivity is given by $\sigma(\omega,T) = F(\omega,T,u)$, where $u$ represents the strength of the irrelevant couplings.
By definition, the RG flow must satisfy
\begin{equation}
s^{1+\delta} F(sT, s \omega,  u(s)) = F(T,\omega,u),
\end{equation}
where $u(s)$ is the coupling that results from $u$ upon an RG flow in which time is rescaled by $s^{-1}$, and where $\delta$ is some as yet unspecified scaling exponent. (For a Fermi liquid, $\delta = 0$, since for $u=0$ the conductivity is a delta function of frequency).
 Now let us choose $s = T_0/T$. This gives
\begin{align}
F(T,\omega,u) &= (T_0/T)^{1+\delta} F(T_0, \omega T_0/T, u(T_0/T))
\end{align}
Therefore, we have at low temperatures and frequencies, taking into account that $u(s) \to 0$ as $s \to \infty$:
\begin{equation}
F(T,\omega,u) \approx T^{-1-\delta} \Sigma(\omega/T),
\end{equation}
where we defined $\Sigma(x) = T_0^{1+\delta} F(T_0, T_0 x, 0)$. Comparing with \eqnref{eq:conductivity_scaling} shows that $\delta = 0$ and $\Sigma(0)$ is a nonzero finite number. But then since the DC conductivity at nonzero temperature and with the irrelevant terms set to zero ($u=0$) is given by
\begin{equation}
\sigma_{\mathrm{DC}} = F(T,0,0) = T^{-1} \Sigma(0),
\end{equation}
which is nonzero and finite, we conclude that the resistivity is intrinsic.
%This means that we are now free to disregard irrelevant terms and consider only the fixed point theory, which we will do for the remainder of this paper.

\subsubsection{More precise argument}
\label{subsec:intrinsic_precise}

 Now we will restate the above argument in a more mathematically precise way.

First of all, let us reformulate Assumption \ref{assump:scaling} from the main text (scaling of conductivity) more precisely. Define the function
\begin{equation}
\Sigma_T(x) := T \sigma(xT, T).
\end{equation}
Then the precise statment of Assumption 2 is that
\begin{equation}
\label{eq:sigma_T_limit}
\lim_{T \to 0_+} \Sigma_T = \Sigma,
\end{equation}
where $\Sigma$ is a continuous function such that $0 < \Sigma(0) < \infty$, and we define the limit of functions in the distribution sense, that is \eqnref{eq:sigma_T_limit} is equivalent to saying that
\begin{equation}
\lim_{T \to 0^+} \int \Sigma_T(x) \varphi(x) dx = \int \Sigma(x) \varphi(x) dx,
\end{equation}
for any compactly supported continuous test function $\varphi$.

Now let us suppose that $\sigma(\omega,T) = F(\omega,T,u_0)$, for some fixed value of the irrelevant couplings $u_0$, where the function $F$ obeys the scaling relation
\begin{equation}
F(\omega, T, u) = s^{1+\delta} F(s\omega, sT, \mathcal{L}_s(u)),
\end{equation}
where $u \mapsto \mathcal{L}_s(u)$ represents the RG flow on $u$.
This implies that, upon choosing some fixed $T_0$:
\begin{align}
\sigma(\omega,T) &= (T_0/T)^{1+\delta} F(T_0, \omega T_0/T, \mathcal{L}_{T_0/T}(u_0)) \\
&= T^{-1-\delta} \widetilde{\Sigma}_{\mathcal{L}_{T_0/T}(u_0)}(\omega/T), \label{eq:sydney}
\end{align}
where we defined $\widetilde{\Sigma}_{u}(x) := T_0^{1+\delta} F(T_0, T_0 x, u)$. Now the definition of an irrelevant coupling implies that $\lim_{s \to \infty} \mathcal{L}_s(u_0) = 0$, and we furthermore assume that $\widetilde{\Sigma}_u(x)$ is continuous in $u$ as $u \to 0$, i.e. $\mathrm{lim}_{u \to 0} \widetilde{\Sigma}_u = \widetilde{\Sigma}_0$ in the distribution sense. [In principle, $\widetilde{\Sigma}_0$ could be a proper distribution, e.g.\ it could include delta function peaks.] Then, from \eqnref{eq:sydney} and the definition of $\Sigma_T(x)$ [no tilde] above, we find that $\Sigma_{T}(x) = T^{-\delta} \widetilde{\Sigma}_{\mathcal{L}_{T_0/T}(u_0)}(x)$.

By taking the limit in the distribution sense, we now obtain the following results depending on $\delta$:
\begin{itemize}
\item If $\delta > 0$, then from $\widetilde{\Sigma}_{\mathcal{L}_{T_0/T}(u)} = T^{\delta} \Sigma_T$ and taking the limit in the distribution sense as $T \to 0^{+}$ we obtain $\widetilde{\Sigma}_0 = 0$ (in the distribution sense, that is, it gives zero when integrated against any test function).
\item If $\delta < 0$, then similarly to above we obtain $\Sigma = 0$ (in the distribution sense). However, this contradicts our assumption that $\Sigma(0) \neq 0$ and $\Sigma$ is continuous, because any such function is not zero in the distribution sense.
\item $\delta = 0$. Then we obtain $\widetilde{\Sigma}_0 = \Sigma$.
\end{itemize}

Observe that the intrinsic conductivity (i.e.\ the conductivity of the IR theory with all irrelevant terms set to zero) is given by $\sigma_{\mathrm{intrinsic}}(\omega, T) = T^{-1-\delta} \widetilde{\Sigma}_0(\omega/T)$. Therefore, the first case of $\delta > 0$, since we found that $\widetilde{\Sigma_0} = 0$ we would conclude that the intrinsic conductivity is strictly zero at all frequencies. This does not seem plausible in a theory with charged degrees of freedom, and we therefore exclude this possibility. Therefore, the only remaining possibility is that $\delta = 0$ and $\widetilde{\Sigma} = \Sigma_0$, which [since by assumption $\Sigma_0(0) < \infty$] implies that the DC resistivity is intrinsic.

\section{The effect of conserved quantities on conductivity}
\label{appendix:susceptibilities}

In this appendix, we derive the formula \eqnref{eq:sigmaomega} for the weight of the delta function in the conductivity $\sigma(\omega)$ at $\omega=0$ for an ersatz Fermi liquid and extend to the case without continuous rotational symmetry.

Suppose we consider a Hamiltonian $H$ that has $k$ extensive conserved Hermitian operators $M_1, \cdots, M_k$. Then we can introduce the corresponding thermodynamically conjugate parameters $\eta^1, \cdots, \eta^k$ and then consider the Gibbs ensemble
\begin{equation}
\label{eq:Gibbs}
\rho = \frac{1}{\mathcal{Z}} e^{-\beta(H - \eta^a M_a)}.
\end{equation}
We will assume that, before being disturbed, the state of the system is just given by the grand canonical ensemble
\begin{equation}
\label{eq:grand_canonical}
\rho = \frac{1}{\mathcal{Z}} e^{-\beta(H - \mu Q)},
\end{equation}
where $Q$ is the total charge operator. Since $Q$ is itself a conserved quantity, it must be expressible as a linear combination of the $M^a$'s, i.e. $Q = q^a M_a$ for some coefficients $q^a$. Thus, \eqnref{eq:grand_canonical} amounts to saying that before being disturbed, $\eta^a = \eta^a_*$, with
\begin{equation}
\label{eq:eta_star}
\eta^a_* = q^a \mu.
\end{equation}

We can define the susceptibility matrix
\begin{equation}
\chi_{ab} := \beta \frac{\partial^2}{\partial \eta^a \partial \eta^b}  \log \mathcal{Z} \biggr|_{\eta = \eta_*}
\end{equation}
Thermodynamic stability requires that this matrix be positive-definite.

We can also define the generalized susceptibility between any two operators as
\begin{equation}
\chi_{AB} := \int_0^\beta \langle A^{\dagger} B(i\lambda) \rangle d\lambda - \beta \langle A^{\dagger} \rangle \langle B \rangle,
\end{equation}
where the expectation values are taken with respect to the ensemble \eqnref{eq:Gibbs}, and $B(i\lambda) = e^{-\lambda K} B e^{\lambda K}$, with $K = H - \mu Q$.
We observe that it is symmetric: $\chi_{AB} = \chi_{BA}^*$.
We can also write
\begin{equation}
\chi_{AB} = \frac{d}{ds} \langle A^\dagger \rangle_{K- sB} \bigg|_{s=0},
\end{equation}
where $\langle \cdot \rangle_{\Gamma}$ denotes an  expectation value taken with respect to $e^{-\beta \Gamma} / (\mathrm{Tr} \, e^{-\beta \Gamma} )$. In particular we have that $\chi_{ab} = \chi_{M^a M^b}$.

Now we consider the case where we set $B = J^i$, the current operator. Then we know that we can write $J^i = i[H,\Pi^i]$, where $\Pi^i$ is the polarization operator. Hence, we find
\begin{align}
\chi_{AJ^i} &= i \int_0^\beta \langle A^\dagger [H, \Pi^i(i\lambda)] \rangle d\lambda \\
&= i \int_0^\beta \langle A^\dagger [H - \mu Q, \Pi^i(i\lambda)] \rangle \\
&= i\left\langle A^\dagger \int_0^\beta \frac{d}{d\lambda} \Pi^i(i\lambda) \right \rangle \\
&= i \langle A^\dagger \Pi^i(i\beta) - A^\dagger \Pi^i \rangle \\
&= i \langle [\Pi^i, A^\dagger] \rangle.
\end{align}
where we used the fact that $Q$ commutes with $\Pi^i$.

Suppose we furthermore assume that $A$ commutes with the Hamiltonian $H$. Then from linear response theory, we know if we switch on an electric field $\mathbf{E}$ (corresponding to replacing the Hamiltonian with $H + E_i \Pi^i$ -- we use the repeated index summation convention), then the rate of change of the expectation value of $A^\dagger$ is
\begin{equation}
\frac{d}{dt} \langle A^\dagger \rangle_\mathbf{E} = iE_i \langle [\Pi^i,A^\dagger] \rangle = E_i \chi_{A J^i}. \label{eq:Adot}
\end{equation}
For example, if $A = p_j = P_j/V$, i.e.\ the momentum density (where $V$ is the total system volume), then we know that in response to an electric field we must have
\begin{equation}
\langle \dot{p_j} \rangle_\mathbf{E} = \mathcal{Q} E_j,
\end{equation}
where $\mathcal{Q}$ is the charge density. Therefore we conclude that
\begin{equation}
\label{eq:pJ_susceptibility}
\chi_{p_j J^i} = \mathcal{Q} \delta\indices{^i_j}.
\end{equation}

Now let us compute the weight of the delta function in the conductivity $\sigma(\omega)$ at $\omega = 0$. Observe that this is equal to $\pi \mathrm{lim}_{t \to \infty} \sigma(t)$, where $\sigma(t)$ is the real-time conductivity. Recall that $\mathcal{E}_i \sigma^{ij}(t)$ describes the current at time $t$ after an electric field impulse $\mathcal{E}_i$ is applied at $t=0$. The reason why $\sigma(t)$ might not go to zero as $t \to \infty$ is that the electric field can push the system into an equilibrium configuration with a different expectation value of the conserved quantities $M^a$, or equivalently a different value of the thermodynamic potentials $\eta_a$. In linear response, $\eta_a$ can only be shifted by an infinitesimal value $\delta \eta_a$, and so we can compute
\begin{align}
J^i(t = \infty)  &= (\delta \eta^a)  \chi_{J^i M_a} \\ &= (\chi^{-1})^{ab} \delta \langle M_b \rangle \chi_{J^i M_a}.
\end{align}
Since we know from \eqnref{eq:Adot} that $\delta 
\langle M_b \rangle = \chi_{J^i M_b} \mathcal{E}_i$, we find that the conductivity tensor is given by
\begin{equation}
\label{eq:conductivity_formula}
\sigma(t = \infty)^{ij} = \frac{1}{V} (\chi^{-1})^{ab} \chi_{J^i M_a} \chi_{J^j M_b}.
\end{equation}
In this derivation we have assumed for simplicity that the energy decouples from the other conserved quantities in the sense that $\chi_{H M^a} = 0$. We show in section \ref{subsec:appendix_to_the_appendix} that the main conclusions of this section are unchanged if one lifts this assumption.

Now let us specialize to the case of an ersatz Fermi liquid in spatial dimension $d=2$, which has the conserved quantities $\hat{n}(\theta)$. From the 't Hooft anomaly of the emergent symmetry group one finds \cite{Else_2007} that in response to an electric field $\mathbf{E}$,
\begin{equation}
\left\langle \frac{d}{dt} \hat{n}(\theta) \right \rangle_{\mathbf{E}} = \frac{V mq}{(2\pi)^2} \epsilon^{i j} E_i \frac{d k_j(\theta)}{d\theta}
\end{equation}
Hence, we find
\begin{equation}
\label{eq:efl_susceptibility}
\chi_{J^i n(\theta)} = \frac{Vmq}{(2\pi)^2} \epsilon^{ij} \frac{dk_j(\theta)}{d\theta},
\end{equation}
where $\mathbf{k}(\theta)$ is the momentum of the point on the Fermi surface specified by $\theta$.

Let us now further specialize to the case where we impose continuous rotational symmetry. In this case we have $\mathbf{k}(\theta) = k_F (\cos \theta, \sin \theta)$.
It is convenient to work in a different basis of the conserved quantities, namely the Fourier components \footnote{Since these are not Hermitian, strictly speaking some of the formulas above must be slightly modified}
\begin{equation}
\hat{n}_l = \frac{1}{2\pi} \int_0^{2\pi} e^{-il\theta} \hat{n}(\theta) d\theta.
\end{equation}
In this basis the susceptibilty matrix must be diagonal:
\begin{equation}
\chi_{l l'} := \chi_{\hat{n}_l \hat{n}_{l'}} = \chi_l \delta_{l,l'},
\end{equation}
while the $x$ component of \eqnref{eq:efl_susceptibility} can be written as
\begin{align}
\chi_{J^x \hat{n}_{l'}} &= \frac{V mq k_F}{8\pi^2} (\delta_{l,1} + \delta_{l,-1})
\end{align}
Substituting into \eqnref{eq:conductivity_formula}, we find
\begin{equation}
\label{eq:sigmachi_1}
\sigma(t = \infty) = V \frac{(mqk_F)^2} {32\pi^4}\frac{1}{\chi_1} 
\end{equation}
from which we can recover \eqnref{eq:sigmaomega} in the main text if we use $P_x = \pi k_F (\hat{n}_1 + \hat{n}_{-1})$ and invoke Luttinger's theorem for ersatz Fermi liquids \cite{Else_2007} to identify the charge density as $\mathcal{Q} = mq k_F^2/(4\pi)$.

Returning to the general case, if we look at \eqnref{eq:conductivity_formula}, then then there are only two ways to get the conductivity tensor $\sigma(t=\infty)$ to zero: either (i) $\chi_{J^i M_a} = 0$ for all $i$, $a$; or else (ii) $\chi^{-1}$ is not positive-definite, i.e.\ it has a zero eigenvalue, corresponding to an infinite eigenvalue of $\chi$.

Possibility (i) turns out to be untenable for describing strange metals as ersatz Fermi liquids. For example, in \eqnref{eq:sigmachi_1} it would correspond to setting $mqk_F = 0$, which would imply that $\mathcal{Q} = 0$. More generally, in any ersatz Fermi liquid in $d=2$ (not necessarily assuming rotational symmetry), then from \eqnref{eq:efl_susceptibility} and the Luttinger's theorem for ersatz Fermi liquids \cite{Else_2007} we conclude that $\chi_{J^i n(\theta)} = 0$ for all $\theta$ would correspond to the charge per unit cell $\nu$ being 0 mod 1, in contradiction to our Assumption \ref{assump:compressible}. One also has a similar conclusion in other spatial dimensions $d$.

Finally, let us prove the claim we made in the main text, that the spatial and time-reversal symmetry properties of the operator whose susceptibility diverges are inherited from those of the current operator.  To see this, we assume that the action of spatial and time-reversal symmetries on the $M_a$'s is described by a real linear representation $\Gamma\indices{_a^b}(g)$, and that there is a positive-definite ``metric tensor'' $\eta_{ab}$ such that $\Gamma$ is metric-preserving
(one can readily verify that this is the case for ersatz Fermi liquids). We introduce operators $O^i$ according to
\begin{equation}
\label{eq:the_O_operator}
O^i = \eta^{ab} \chi_{J^i M_a} M_b.
\end{equation}
Clearly, $O^i$ inherits the symmetry transformation properties of the current $J^i$. What we want to prove is that the susceptibility of $O^i$ must diverge.

We can work in a basis in which $\eta$ is the identity matrix. 
Since $\chi^{-1}$ is symmetric, it follows that there exists an orthogonal matrix $U$ such that $U \chi^{-1} U^{-1} = D$, where $D$ is a diagonal matrix whose diagonal entries are the eigenvalues of $\chi^{-1}$.
 In order for the weight \eqnref{eq:conductivity_formula} to be fully suppressed, it is necessary for some of the eigenvalues of $\chi^{-1}$ to be zero.
 To avoid issues with inverting a singular matrix, we will instead consider what happens as certain eigenvalues of $\chi^{-1}$ \emph{approach} zero as a parameter $\gamma$ of the Hamiltonian goes to zero. Then $U$ and $D$ are functions of $\gamma$. However, as the space of orthogonal matrices is compact, one is entitled to assume that $U_\gamma$ approaches a limit $U_*$ as $\gamma \to 0$.  Since we are only interested in the asymptotic behavior as $\gamma \to 0$, we can assume that $U_\gamma = U_*$ independently of $\gamma$. Now we work in the basis described by the change of basis matrix $U_*$, so that $\chi$ and $\chi^{-1}$ are diagonal, specifically:
 \begin{equation}
 \chi_{a b} = \chi_a \delta_{a b}
 \end{equation}
 (no implicit summation here and in the following).
For some set $\mathcal{I}$, we have that $\lim_{\gamma \to 0} \chi_a^{-1} = 0$ for $a \in \mathcal{I}$.
Then we have
\begin{align}
\label{eq:O_susceptibility}
\chi_{O^i O^i}
&= \sum_a (\chi_{M_\alpha J^i})^2 \chi_a,
\end{align}
and [from \eqnref{eq:conductivity_formula}]:
\begin{align}
\label{eq:a_summation}
\sigma(t=\infty)^{ii}
&= \sum_a (\chi_{M_a J^i})^2 \chi_a^{-1}.
\end{align}
 $\chi_{M_a J^i}$ must be nonzero for at least one $a \in \mathcal{I}$, otherwise from \eqnref{eq:a_summation} the divergence of $\chi_a$ would not suppress $\sigma(t=\infty)^{ii}$. But then from \eqnref{eq:O_susceptibility}, we conclude that $\chi_{O^i O^i}$ diverges.
 
\subsubsection{Effect of mixing with the energy}
\label{subsec:appendix_to_the_appendix}
Here we discuss how the results above are modified in the case where $\chi_{H M^a} \neq 0$. The idea is to define $M^0 = H$, and make indices range from $0$ to $k$ instead of $1$ to $k$. Then most of the formulas above will still hold with the exception of \eqnref{eq:Gibbs}, which should be rewritten as
\begin{equation}
\rho = \frac{1}{\mathcal{Z}} e^{-\beta \eta^a H_a},
\end{equation}
and \eqnref{eq:eta_star}, which becomes
\begin{align}
\eta_*^0 &= 1, \\
\eta_*^a &= q^a \mu \quad( a \in \{ 1, \cdots, k \}).
\end{align}

We then see that suppressing the weight of the delta function will require that the $(k+1)\times(k+1)$ matrix $(\chi^{-1})^{a b}$ has a zero eigenvalue. Furthermore, we remark that the operator $O^i$ defined in \eqnref{eq:the_O_operator}, whose susceptibility diverges, never actually involves the Hamiltonian, since $\chi_{J^i H}$ can be argued to be zero by invoking Bloch's theorem that states that the expectation value of the current density in an equilibrium state is zero \cite{Watanabe_1904}.

\bibliography{ref-autobib,ref-manual}
%\printbibliography[segment=1]

\end{document}